\documentclass[%
 reprint,
 superscriptaddress,
 amsmath,amssymb,
 aps,
]{revtex4-1}

\usepackage{graphicx}
\usepackage{dcolumn}
\usepackage{bm}
\usepackage{color}

\begin{document}

\preprint{APS/123-QED}

\title{Terahertz Wave Guiding by Femtosecond Laser Filament in Air}

\author{Jiayu Zhao}
\affiliation{%
Institute of Modern Optics, Nankai University, Key Laboratory of Optical Information Science and Technology, Ministry of Education, Tianjin 300071, China
}

\author{Yizhu Zhang}
\affiliation{%
Shanghai Advanced Research Institute, Chinese Academy of Sciences, Shanghai 201210, China
}

\author{Zhi Wang}
\email{zhiwang@nankai.edu.cn}
\affiliation{%
Institute of Modern Optics, Nankai University, Key Laboratory of Optical Information Science and Technology, Ministry of Education, Tianjin 300071, China
}

\author{Wei Chu}
\author{Bin Zeng}
\author{Ya Cheng}
\author{Zhizhan Xu}
\affiliation{%
State Key Laboratory of High Field Laser Physics, Shanghai Institute of Optics and Fine Mechanics, Chinese Academy of Sciences, Shanghai 201800, China
}

\author{Weiwei Liu}
\email{liuweiwei@nankai.edu.cn}
\affiliation{%
Institute of Modern Optics, Nankai University, Key Laboratory of Optical Information Science and Technology, Ministry of Education, Tianjin 300071, China
}

\date{\today}

\begin{abstract}
Femtosecond laser filament generates strong terahertz (THz) pulse in air. In this paper, THz pulse waveform generated by femtosecond laser filament has been experimentally investigated as a function of the length of the filament. Superluminal propagation of THz pulse has been uncovered, indicating that the filament creates a THz waveguide in air. Numerical simulation has confirmed that the waveguide is formed because of the radially non-uniform refractive index distribution inside the filament. The underlying physical mechanisms and the control techniques of this type THz pulse generation method might be revisited based on our findings. It might also potentially open a new approach for long-distance propagation of THz wave in air.
\end{abstract}

\pacs{Valid PACS appear here}
\maketitle


Strong diffraction and energy attenuation due to water vapor absorption are major obstacles to expedite the application of terahertz (THz) technology in air. Because of the capability to generate THz wave at a remote distance up to hundreds of meters,  femtosecond laser filamentation, a unique nonlinear optical phenomenon being able to generate a long plasma channel in air, has been suggested to provide a solution to this problem [1]. And the plasma channel is often referred to as a ``filament''. For example, THz remote sensing in atmosphere by using filamentation has been demonstrated to be realistic recently [2]. This cutting edge THz wave generation technology has also been highly expected in applications such as determining the carrier envelop phase of few-cycle pulse [3], studying the ionization dynamic [4], ultra-broadband THz detection [5, 6] and spectroscopy [7], etc.

However, the fundamental nature of the THz wave propagation during the filamentation is barely known. Experiments have indicated that the THz pulse generated by the filament is highly directional. It has been explained by the off-axis phase matching [8, 9]. In our current work, we have observed superluminal propagation of THz pulse during filamentation process. It provides strong evidence that the THz pulse is guided along the filament. Simulation results have confirmed that non-uniform transverse distribution of the plasma density inside the filament will give rise to the formation of photonic waveguide at THz band. THz energy is strongly confined into a sub-wavelength scale. The uncovered new aspect would be an important concern not only for studying the underlying physical mechanism, but also for exploiting this advanced THz wave generation and propagation technology.

The experimental setup adapts the typical THz generation and detection scheme by using filamentation in air [10]. The laser beam was Gaussian and the diameter $d$ = 1 cm (1/$e^{2}$). In brief, a 1 kHz, 800 nm, 50 fs Ti: sapphire laser pulse was split into two paths. One was the pump beam and the other was used as the probe of time domain spectroscopy (TDS) detection [10]. The energy of the pump beam was about 1 mJ/pulse and focused by a $f$ = 110 cm lens, creating a centimeter-scale long filament in air. The exiting THz pulse from the plasma was first collimated by an off-axis parabolic mirror ($D$ = 50 mm, $f$ = 100 mm), then focused by another identical parabolic mirror onto a 1.5 mm-thick ZnTe crystal. The probe beam was combined with THz pulse by a Pellicle beam splitter, performing TDS measurement. Particularly, a Teflon plate, which has high transmission for THz, was put inside the filament at different distances during our experiment. As sketched in Fig. 1(a), it not only blocked the transmission of fundamental 800 nm light, but also interrupted the formation of plasma filament. No penetrating-through hole was found on the Teflon plate after the experiment.
\begin{figure}[htb]
\begin{center}
\includegraphics[scale=0.35]{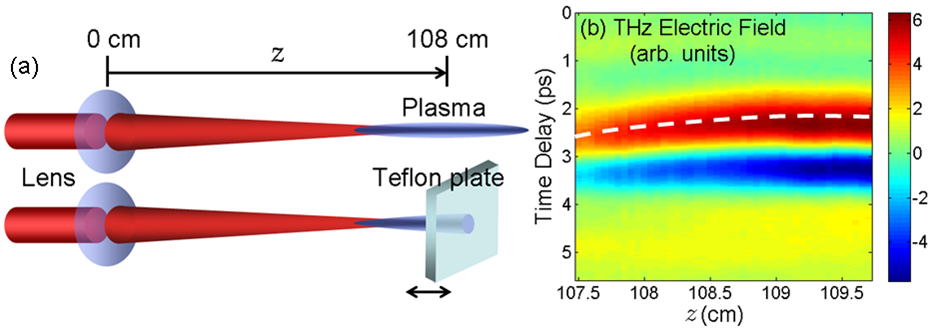}
\caption{\label{fig:epsart} (Color online) (a) A Teflon plate, which can be moved by a motor stage along the laser propagation direction in a step length of 0.5 mm, was inserted into the plasma column. (b) The recorded THz waveforms as a function of propagation distance. The dashed white line highlights the superluminal phenomenon.}
\end{center}
\end{figure}

THz waveforms recorded as a function of the propagation distance $z$ are presented in Fig. 1(b), which are essentially plotted in the time coordinate moving at group velocity of the probe beam (800 nm pulse). The propagation distances correspond to the inserting positions of the Teflon plate and $z$ = 0 is identified as the position of the focusing lens. It could be seen that THz pulse retains the characteristic of single cycle. Fig. 2(a) outlines the evolution of the THz peak-to-peak amplitude (red triangles) as a function of the propagation distance. It continues increasing from $z$ = 107.5 cm, where the detection system starts to be able to detect the THz pulse. However, the peak-to-peak amplitude tends to become saturated after $z$ = 109.5 cm, which is almost the ending distance of the filament.

However, what impresses us mostly is that the maximum of the THz waveform moves significantly forward as the pulse propagates further. This trend is highlighted in Fig. 1(b) by a dashed white line which indicates the temporal trajectory of the amplitude maximum. Fig. 2(a) further quantifies the displacement of the THz waveform maximum (blue circles). The total temporal advance from $z$ = 107.5 cm to $z$ = 108.5 cm is about 0.3 ps. It is noticed that this region superposes the zone where significant nitrogen ($N_2$) fluorescence signal is produced (see Fig. 2(a) black squares). The longitudinal distribution of $N_2$ fluorescence signal, which essentially corresponds to the longitudinal free electron density evolution, was obtained by a side-imaging approach [11]. Beyond this region, no significant superluminal propagation could be observed.
\begin{figure}[htb]
\begin{center}
\includegraphics[scale=0.3]{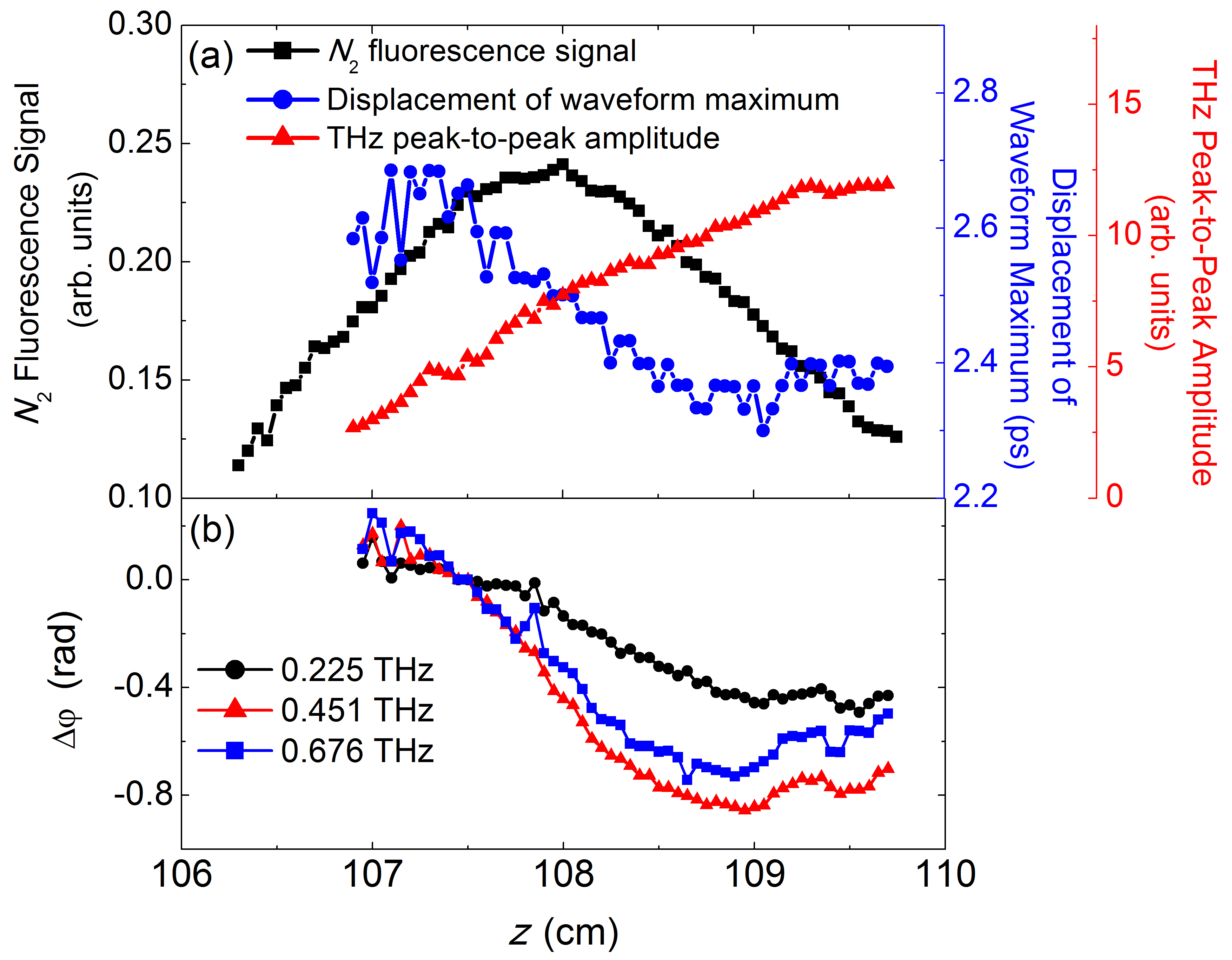}
\caption{\label{fig:epsart} (Color online) (a) THz peak-to-peak amplitude (red triangles), displacement of the THz amplitude maximum (blue circles), $N_2$ fluorescence signal of plasma filament (black squares) and (b) Phase variations for three typical frequencies: as a function of the propagation distance.}
\end{center}
\end{figure}

In order to quantitatively investigate the superluminal propagation of the THz pulse, fast-Fourier-transformations (FFT) on each waveform presented in Fig. 1(b) have been performed. In this way, the phases and amplitudes of different frequencies could be retrieved. The resolved phase variations as a function of propagation distance are illustrated in Fig. 2(b) for three typical frequencies, namely 0.225 THz, 0.451 THz and 0.676 THz, as black circles, red triangles and blue squares, respectively. The phases of three frequencies move forward within the regime from $z$ = 107.5 cm to $z$ = 108.5 cm, indicating superluminal propagations. After $z$ = 108.5 cm, the phases tend to be constant. The phase variation ${\Delta\varphi}$ is in fact given by:
\begin{eqnarray}
\Delta \varphi = \left ( n_{THz}-n_{g,800~nm} \right )\frac{\Omega }{c}\Delta z,
\end{eqnarray}
where $n_{THz}$ denotes the refractive index of THz wave, $n_{g,800~nm}$ is referred to as the group index of the probe beam, $c$ is the light speed in vacuum and ${\Omega,\Delta z}$ indicate the THz wave frequency and propagation distance, respectively. According to Eq. (1), the refractive index of THz frequencies could be estimated by the slopes of the phase variation curves shown in Fig. 2(b) between $z$ = 107.5 cm and $z$ = 108.5 cm. The results give rise to $n_{0.225~ THz}$ = 0.99244, $n_{0.451~THz}$ = 0.99354 and $n_{0.676~THz}$ = 0.99603. The same analysis procedures are also carried out for other frequencies. The obtained refractive indices are shown in Fig. 3 as black squares. Here, we have taken $n_{g,800~nm}$ = 1.00028 calculated on the basis of Sellmeier formula in air. Note that the refractive index of THz wave in air $n_{THz,air}$ is slightly higher than 1. The significant discrepancy between $n_{THz}$ and $n_{THz,air}$ implies that the detected THz pulse does not propagate in air as expected. The observed superluminal phenomenon could only be given rise by the self-guided propagation inside the filament. It is worth mentioning that since it is well known that a gain may induce the superluminal phenomenon of the group velocity of a pulse, we focus on the study of the phase velocity in the current work to confirm the self-guided propagation of THz pulse inside the filament.
\begin{figure}[h]
\begin{center}
\includegraphics[scale=0.26]{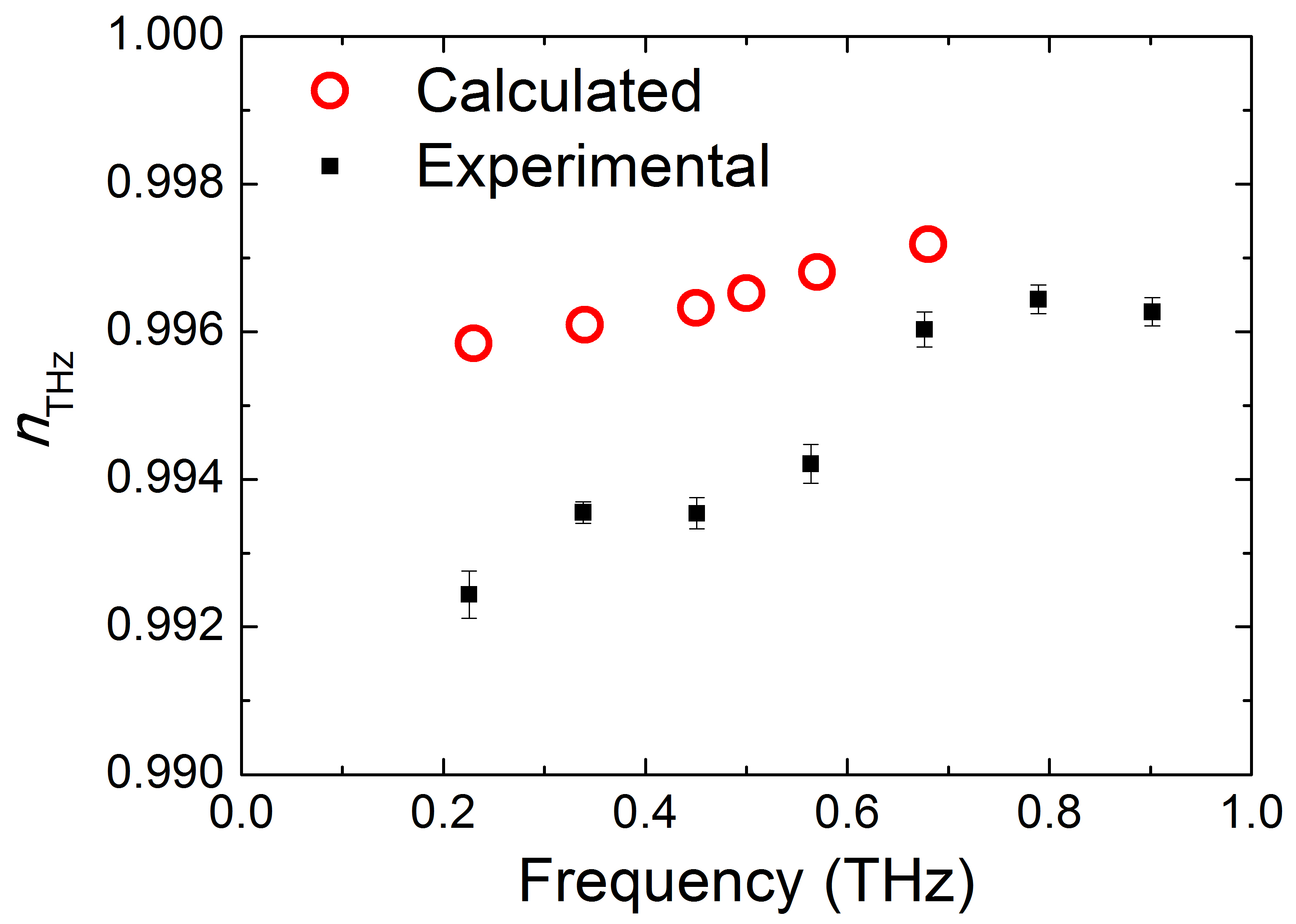}
\caption{\label{fig:epsart} (Color online) The retrieved refractive indices from experiment (black squares) and numerical simulations (red circles) for different frequencies. Numerical simulations are performed by COMSOL in order to obtain the effective refractive indices of the interested propagation modes shown in Fig. 6.}
\end{center}
\end{figure}

When propagating inside plasma, the refractive index of THz wave is the real part of the square root of the permittivity:
\begin{eqnarray}
n_{THz}=Re\left ( \sqrt{1-\frac{\omega _{p}^{2}}{\Omega ^{2}-i\nu \Omega }} \right ),
\end{eqnarray}
where ${\omega _{p}}$ indicates plasma frequency (in SI units):
\begin{eqnarray}
\omega _{p}=\sqrt{\frac{e^{2}}{m_{e}\varepsilon_{0}}N_{e}}.
\end{eqnarray}
$N_e$ denotes the number density of electrons, $e$ represents the electric charge, ${m_e}$ indicates the effective mass of the electron and ${\varepsilon_{0}}$ is the permittivity in vacuum. In Eq. (2), $\nu$ $\sim$ 1 THz corresponds to the typical electron collision frequency inside filament [12]. Note in Eq.(2) and (3), $N_e$ is the only undetermined variable.

Therefore, we have performed numerical simulations to infer the plasma density inside the filament. It is based on the nonlinear wave equation using slowly varying envelop approximation written in the retarded coordinate system $\tau =t-z/\upsilon _{g}(\omega )$ as [13]:
\begin{eqnarray}
&2ik_{0}\frac{\partial A}{\partial z}+\Delta _{\perp }A+2\left ( 1+\frac{i}{\omega } \frac{\partial }{\partial t}\right )\frac{k_{0}^{2}}{n_{0}}\left ( \Delta n_{Kerr}+\Delta n_{plasma} \right )A\nonumber& \nonumber \\
&-k_{2}k_{0}\frac{\partial ^{2}A}{\partial \tau ^{2}}-ik_{0}\alpha A=0.&
\end{eqnarray}
Eq. (4) involves a number of optical effects such as diffraction, self-focusing, group-velocity dispersion, self-steepening as well as plasma generation and energy losses due to multi-photon/tunnel ionization. Here, $A$ is the electric field envelope function. $\upsilon _{g}\left ( \omega  \right ), k_{0}, k_{2}$ and $\alpha $ represent group velocity, wave number, group velocity dispersion parameter and absorption coefficiency associated with ionization in air, respectively [13]. Similar model has been successfully used in [14]. The obtained plasma density inside a filament is in good agreement with the experimental results [14]. The modeling of the plasma generation is the same as Ref. [13]. Note that in our experiment, the pumping pulse's duration is much shorter than the THz pulse. The interaction between them might not be significant and is not taken into account in our model.

The experimental parameters were adapted as the initial parameter of the simulations. The simulation was started after the laser pulse has passed through the focusing lens. The obtained $N_e(r, z)$ by numerical simulation is shown in Fig. 4(a). And the on-axis peak electron density $N_e (0, z)_{max}$ of this few-centimeter-long filament reaches about $1.4 \times 10^{17}$ cm$^{-3}$. We have attempted to substitute $N_e = 1.4 \times 10^{17}$ cm$^{-3}$ or $ N_e = 7 \times 10^{16}$ cm$^{-3}$ into Eq. (2) and (3). The calculated on-axis refractive indices (e.g. $n_{0.225~THz}$ = 4.09945, $n_{0.451~THz}$ = 1.23819 and $ n_{0.676~THz}$ = 0.57944) are still far different from our experiment results (black squares in Fig. 3).
\begin{figure}[htb]
\begin{center}
\includegraphics[scale=0.35]{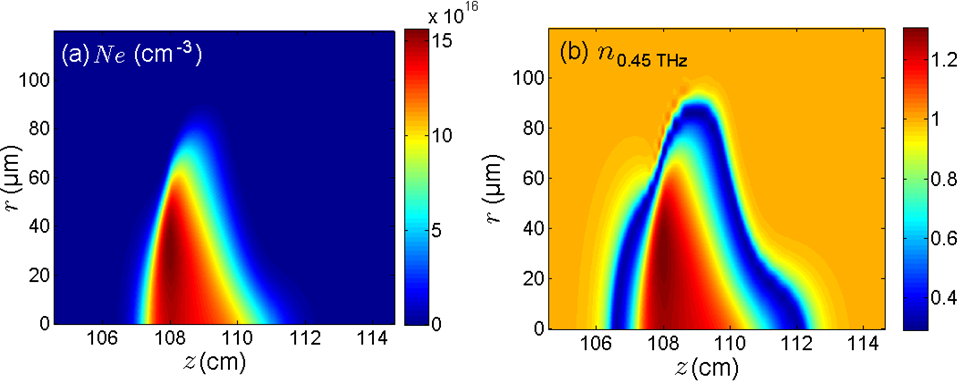}
\caption{\label{fig:epsart} (Color online) (a) Plasma density and (b) Refractive indices distribution of 0.45 THz: inside the plasma filament by numerical simulations according to Eq. (4).}
\end{center}
\end{figure}

However, we have noticed that the corresponding refractive index distribution of Fig. 4(a) features pronounced non-uniformity. Then, according to Eq. (2) and (3), the refractive index at 0.45 THz has been calculated for an example. The result is shown in Fig. 4(b). Particularly, the representative radial index distribution at $z$ = 108 cm is depicted as the thin solid red line in Fig. 5. A refractive index dip appears at the vicinity of $r$ = 70 $\mu$m, where the corresponding plasma density is about $ 7.5 \times 10^{15}$ cm$^{-3}$ as could be seen from the dashed blue line in Fig. 5. A question is then naturally raised up: could the modulated refractive index distribution give rise to photonic wave guiding effect?
\begin{figure}[htb]
\begin{center}
\includegraphics[scale=0.3]{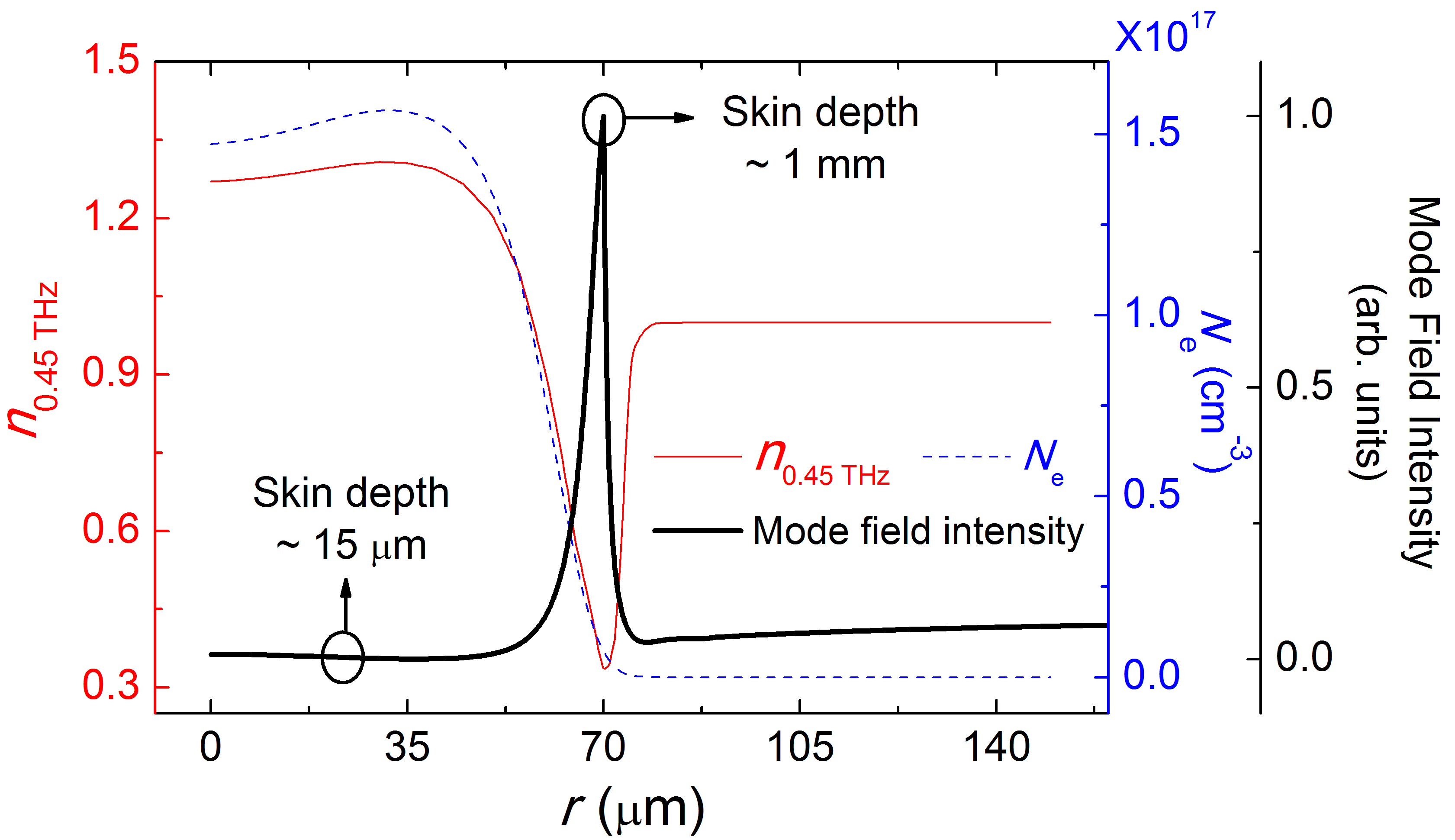}
\caption{\label{fig:epsart} (Color online) Plasma density (dashed blue line), refractive indices (thin solid red line) and mode field intensity (thick solid black line) of 0.45 THz as a function of $r$ at $z$ = 108 cm.}
\end{center}
\end{figure}

In order to get a clue of the mechanism of the THz wave self-guiding, we calculate the eigenmodes around the filament area at $z$ = 108 cm by the full-vector finite-element method (FEM) with the commercial software COMSOL Multiphysics. The refractive index distribution in the plasma filament is accurately considered in our FEM model by using quadratic triangular elements with a maximum size 1 $\mu$m and the perfectly matched layers (PMLs) at the boundary of air area. The doublet degenerated modes localized in the filament area are found in our simulation. Fig. 6 shows the intensity profiles ($z$-component of the Poynting vector) and electric field vector of these modes at 0.23 THz, 0.45 THz and 0.68 THz. Since the fields are identical after a rotation of $\pi /2$ radian, only one of the doublet degenerated modes is illustrated at 0.23 THz and 0.68 THz. As shown in the figure, the intensity fields are localized in a circular region and the electric field vector is nearly radial polarization for these modes. To compare with the refractive index distribution, the mode field intensity at 0.45 THz is also shown in Fig. 5 (thick solid black line) along the white line in Fig. 6(c). At 0.45 THz, the mode fields are strongly confined in a sub-wavelength region, and the maximum light field is located at the refractive index dip. Whereas, at 0.23 THz and 0.68 THz, because the refractive index dip is shallow, the mode field tends to extending to the air area and coupling with the radiation modes. However, the radiation mode may diffract strongly in air and could not be collected efficiently by our experimental setup. The calculated effective refractive indices of these localized modes at different frequencies are shown in the Fig. 3 (red circles). The effective refractive indices are less than unity and increase with increasing frequency. Therefore, these modes could contribute to superluminal propagation of THz wave in the plasma filament. Compared with the experimental results, the calculated refractive indices are slightly higher, which may be attributed to the difference of the refractive index distribution along the plasma channel. It is also interesting to notice that the polarization shown in Fig. 6 agrees with the experimental results reported in Ref. [10] that the polarization of the THz pulse generated by a filament in air could not be characterized as a vortex or linear polarization.

On the other hand, we have also calculated the corresponding skin depth at 0.45 THz by the Eq. (3) in Ref. [15]. Note that the ``skin depth'' is determined by the imaginary part of the complex refractive index and used to define the propagation distance after which flux density will drop by a factor of 1/$e$ when a monochromatic electromagnetic wave propagates inside a conductor (in our case, it is a plasma). As indicated in Fig. 5, in the range of $r$ $<$ 50 $\mu$m, the skin depth is less than 15 $\mu$m, while a sharp growth to 1 mm occurs at $r$ $\sim$ 70 $\mu$m. Hence, it is very likely that THz pulse could propagate on the periphery of the filament instead of inside the filament core.
\begin{figure}[h]
\begin{center}
\includegraphics[scale=0.4]{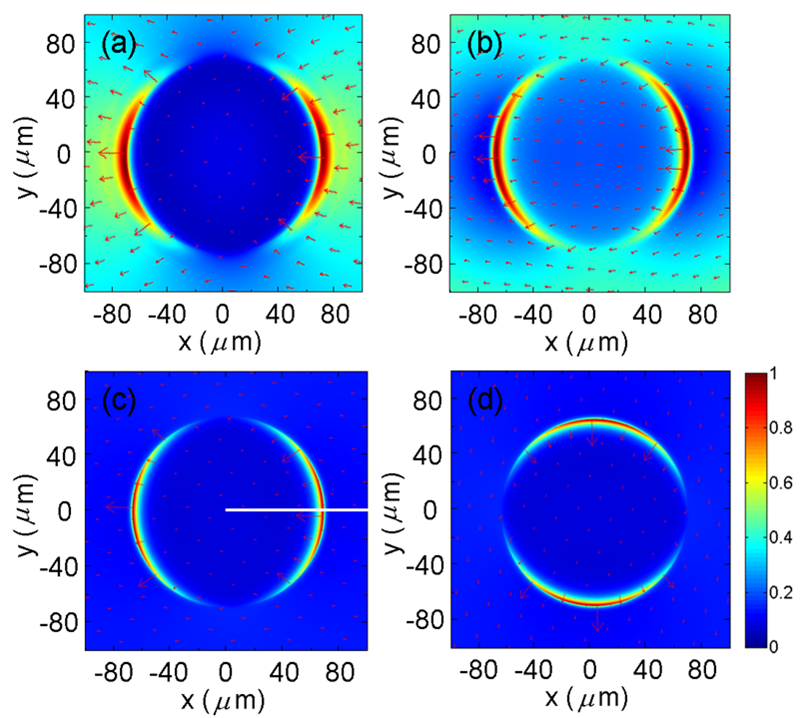}
\caption{\label{fig:epsart} (Color online) $z$-component of the Poynting vector (color) and electric vector (arrows) of the THz eigenmodes around the filament area. (a), (b) one of the doublet degenerated modes at 0.23 THz and 0.68 THz; (c), (d) the doublet degenerated modes at 0.45 THz.}
\end{center}
\end{figure}

In summary, by recording the THz waveform as a function of propagation distance, we have observed superluminal propagating THz pulse produced by the femtosecond laser filament. The superluminal propagation indicates that the detected THz pulse in the forward direction is mainly guided inside plasma channel. Our results have revealed new aspect of the THz generation by femtosecond laser-gas plasma interaction, which is an extensively attractive research topic recently. Not only the underlying physical mechanisms of THz pulse generation, but also the scheme of control techniques might be revisited based on our findings. It might potentially open a new approach to guide THz wave in air.
\\
\\
This work is financially supported by National Basic Research Program of China (2014CB339802, 2011CB808100 and 2013CB922200) and National Natural Science Foundation of China (11174156, 10974213, 60825406, 11274232 and 11174155). WL and ZYZ acknowledge the support of the open research funds of State Key Laboratory of High Field Laser Physics (SIOM).

\nocite{*}
\bibliography{apssamp}

\end{document}